\documentclass{article}
\usepackage{spconf}
\usepackage{cite}
\usepackage{amsmath,amssymb,amsfonts}
\usepackage{algorithm}
\usepackage{algorithmic}
\usepackage{amsmath}
\usepackage{siunitx}
\usepackage{array}
\usepackage{multirow}
\usepackage{graphicx}
\usepackage{threeparttable}
\usepackage{caption}
\usepackage{textcomp}
\usepackage{xcolor}
\usepackage{float} 
\usepackage{subfigure}
\usepackage{subcaption}

\title{ATTACK AND DEFENSE ANALYSIS OF LEARNED IMAGE COMPRESSION}
\name{Tianyu~Zhu$^1$, Heming~Sun$^2$\sthanks{Corresponding author: Heming Sun (sun-heming-vg@ynu.ac.jp)}, Xiankui~Xiong$^3$, Xuanpeng~Zhu$^3$, Yong~Gong$^4$, Minge~jing$^1$, Yibo~Fan$^1$}
\address{$^1$State Key Lab of Integrated Chip System, Fudan University, China\\$^2$Faculty of Engineering, Yokohama National University, Japan\\$^3$State Key Laboratory of Mobile Network and Mobile Multimedia Technology, ZTE Corporation\\$^4$China Nanhu Academy of Electronics and Information Technology}

\begin{document}
\ninept
\maketitle
\begin{abstract}
Learned image compression (LIC) is becoming more and more popular these years with its high efficiency and outstanding compression quality. 
Still, the practicality against modified inputs added with specific noise could not be ignored. 
White-box attacks such as FGSM and PGD use only gradient to compute adversarial images that mislead LIC models to output unexpected results. 
Our experiments compare the effects of different dimensions such as attack methods, models, qualities, and targets, concluding that in the worst case, there is a 61.55\% decrease in PSNR or a 19.15 times increase in bpp under the PGD attack. 
To improve their robustness, we conduct adversarial training by adding adversarial images into the training datasets, which obtains a 95.52\% decrease in the R-D cost of the most vulnerable LIC model. 
We further test the robustness of H.266, whose better performance on reconstruction quality extends its possibility to defend one-step or iterative adversarial attacks.
\end{abstract}
\section{Introduction}
\label{intuoduction}
Image compression acts as the basement of all the other vision tasks. 
For the demands of high efficiency, LIC models have attracted more and more attention. 
However, the forward presentation of neural networks is usually irreversible without quantization error.
This provides a fatal flaw of LIC models that could be fooled by tiny perturbations added to the input images because quantization error would be accumulated through layer-by-layer computation, leading to significant changes in the coding and reconstruction of images.

The adversarial attack is one of the attack methods that utilize the abovementioned disadvantage. 
The attacker intends to modify original images with imperceptible targeted noise for well-pre-trained neural networks to cause undesired behaviors. 
As for LIC models, Chen et al.~\cite{Towards} carried out attacks targeting reconstruction quality by injecting random noise into the original image and optimizing it with a gradient descent algorithm. 
However, comparing the effects of different attack dimensions has not been researched yet.

Meanwhile, such attacks could cause danger to the applications of LIC models. 
For example, security monitoring meets the challenge of distinguishing clues on the screen while local users have to pay more storage resources and transmission bandwidth to obtain images. 
Thus, it’s necessary to explore effective defense methods to weaken these attacks' impact and improve the robustness of common LIC models. 

In this work, we make the following contributions: 
\begin{itemize}
    \item We conduct iterative FGSM and PGD attacks on 6 LIC models of low and high qualities (Sec.~\ref{experiments}). Also, we conduct PGD training~\cite{transfer}\cite{tranferfgsm} as a defense method to finetune the LIC models and raise their robustness (Sec.~\ref{defense method});
    \item We target not only reconstruction quality but also bit rate in attacks and defense, proposing R-D cost change to measure the finetuning effect of defense.
    \item We try a transferring attack on the conventional image compression method by using adversarial images generated by one-step FGSM, which concludes its vulnerability on bit rate (Sec.~\ref{conventional methods}).
\end{itemize}

\section{Related Work}
\label{related work}

\subsection{Learned Image Compression}
\label{learned image comp}
Ball{\'e} et al.~\cite{bmshj2018-fac} first proposed a LIC model in 2016 with factorized-prior as the entropy model (named `\emph{\texttt{Fac}}’), which established the optimization function aiming at the trade-off between distortion and bit rate (R-D cost). 
Minnen et al.~\cite{mbt2018} (named `\emph{\texttt{Mean}}’) and Ball{\'e} et al.~\cite{bmshj2018-hyper} (named `\emph{\texttt{Hyper}}’) applied hierarchical hyper-prior entropy models to compact latent structure information, the latter differs in using only zero-mean Gaussian distribution. 
Another model put forward by Minnen et al.~\cite{mbt2018} (named `\emph{\texttt{Mbt}}’) added an autoregressive context model making entropy estimation even more accurate. 
Based on \texttt{Mbt}, Cheng et al.~\cite{cheng2020} utilized residual blocks in the analysis and synthesis transforms (named `\emph{\texttt{Anchor}}’). 
What’s more, Cheng et al.~\cite{cheng2020} added self-attention modules (named `\emph{\texttt{Attn}}’) to enhance the performance of image compression and reconstruction. 
There are quite a few works about LIC.~\cite{ma3}\cite{lic1}\cite{lic3}\cite{lic4}\cite{ma1}\cite{lic6}\cite{lic5}\cite{ma2}\cite{lic2}

\subsection{Adversarial Attack and Defense}
\label{adv attack&defense}
Most adversarial attacks can be classified as white-box attacks, which means attackers have access to all the information of target networks so that they can take advantage of the weakness.~\cite{survey} 
The goal of a white-box attack is to generate an adversarial image x’ that is similar enough to the original input x but leads to a mistaken result through the target network. 
Given a neural network model F, it can be formulated as:
\begin{equation}
    max\Vert{F(x')}-F(x)\Vert~s.t.~\Vert{x'}-x\Vert\leq\epsilon\label{eq1}
\end{equation}

Here $\Vert\cdot\Vert$ quantitatively describes the difference between two items. $\epsilon$ refers to the maximum perturbation allowed to add to the original input.

Biggio et al.~\cite{firstattack} first reported adversarial images that deceive neural networks and raised attention to the robustness of research. 
Many other attack methods designed for image classification tasks were proposed after that like Fast Gradient Sign Method (FGSM)~\cite{fgsm},  Basic Iterative Method (BIM)~\cite{bim1}\cite{bim2}, Projected Gradient Descent (PGD)~\cite{PGD} and so on. 

The defense method is also a crucial topic of adversarial attack. 
Goodfellow et al.~\cite{fgsm} first suggested using adversarial images generated by FGSM to train the network to learn to handle misleading inputs correctly. 
On their basis, Madry et al.~\cite{PGD} substituted FGSM with PGD so that the finetuned model can improve robustness against most single-step and iterative attacks.

\section{Adversarial Attacks}
\label{attacks}

\subsection{Loss Functions}
\label{loss functions}
Adversarial attacks on LIC models usually focus on reconstruction quality and bit rate, as most LIC models choose R-D cost as their optimization goal.

While targeting reconstruction quality (PSNR attack), based on~\eqref{eq1}, we take peak signal-to-noise ratio (PSNR) as the explanation of $\Vert\cdot\Vert$ so that the attack loss function can be defined as:
\begin{equation}
    L_{PSNR}=L_2~loss(F(x'), x)\label{eq2}
\end{equation} 

Another target is to attack bit rate (bpp attack), which is measured by bits per pixel (bpp). Since the entropy coders of LIC models are specially optimized for high efficiency and accuracy, the entropy of the quantized latent nicely estimates its exact bit length. 
Thus, the attack loss function is shown in~\eqref{eq3}:
\begin{equation}
    L_{bpp}=\frac{\sum\mathbb{E}[-log_2(P(\widehat{y'}))]+\sum\mathbb{E}[-log_2(P(\widehat{z'}))]}{number~of~pixels},\label{eq3}
\end{equation}
where P($\cdot$) means the output of the entropy model. 
As described in~\cite{sun}, for those with a hyper-prior entropy model, $\hat{z}$ is also entropy coded with $\hat{y}$ so that~\eqref{eq3} will include an additional term of $\hat{z'}$.

\subsection{FGSM Attack}
\label{fgsm}
FGSM~\cite{fgsm} is a computationally efficient method for generating adversarial images. 
By considering the sign of the gradients, FGSM determines the direction in which the original image should be modified to create the most significant impact on the model's output in terms of a specific attack target.

\begin{algorithm}
    \caption{FGSM attack}
    \label{fgsm algorithm}
    \begin{algorithmic}[1]
            \REQUIRE original image $X$, max iteration step $T$, step size $\delta$, max perturbation size $\epsilon$
        \ENSURE adversarial image $X'$
        \STATE Let $t=0$, $X'=X$
        \WHILE{$t<T$}
	  \STATE Calculate attack loss L by~\eqref{eq2} or~\eqref{eq3}
	  \STATE Calculate the gradient of the attack loss:\\$grad=\frac{\partial{L(F(X'), X)}}{\partial{X}}$
	  \STATE Calculate and add noise: $X'=X'+\delta{\cdot}sign(grad)$
        \STATE Restrict the total perturbation:
	  \IF{$\Vert{X'}-X\Vert>\epsilon$}
	  \STATE Break
        \ENDIF
	  \STATE $t=t+1$
        \ENDWHILE
        \RETURN $X'$
    \end{algorithmic}
\end{algorithm}

As inspired, our attack adopts an iteration version of FGSM. 
As shown in Algorithm~\ref{fgsm algorithm}, we repeat to add noise calculated by gradients until reaching the max iteration step or upon the maximum allowed perturbation size $\epsilon$. 
The step size $\delta$ is set as \SI{1e-4}{} so that the noise added to the target image is visually unperceivable. 
Meanwhile, we choose $\epsilon=7/255$ in our experiment, which guarantees the PSNR between the adversarial and original image is around 30.

\begin{table*}[]
\tiny
\centering
\renewcommand{\arraystretch}{1.1}
\caption{Summary of bpp, PSNR, and MS-SSIM change of 6 LIC models under FGSM and PGD attack targeting reconstruction quality.}
\label{psnr attack data}
\resizebox{\textwidth}{!}{%
\begin{tabular}{c|c|ccc|ccc}
\hline
\multirow{2}{*}{Method} & \multirow{2}{*}{Model} & \multicolumn{3}{c|}{quality = 2} & \multicolumn{3}{c}{*quality = 8} \\ \cline{3-8} 
 &  & \multicolumn{1}{c|}{bpp change} & \multicolumn{1}{c|}{PSNR change} & MS-SSIM change & \multicolumn{1}{c|}{bpp change} & \multicolumn{1}{c|}{PSNR change} & MS-SSIM change \\ \hline
\multirow{6}{*}{\textbf{FGSM}} & \texttt{Fac} & \multicolumn{1}{c|}{0.88} & \multicolumn{1}{c|}{-11.45\%} & -7.89\% & \multicolumn{1}{c|}{0.99} & \multicolumn{1}{c|}{-26.11\%} & -7.98\% \\
 & \texttt{Mean} & \multicolumn{1}{c|}{0.62} & \multicolumn{1}{c|}{-12.17\%} & -12.37\% & \multicolumn{1}{c|}{1.03} & \multicolumn{1}{c|}{-27.73\%} & -8.71\% \\
 & \texttt{Hyper} & \multicolumn{1}{c|}{0.67} & \multicolumn{1}{c|}{-14.30\%} & -16.90\% & \multicolumn{1}{c|}{1.01} & \multicolumn{1}{c|}{-26.72\%} & -7.41\% \\
 & \texttt{Mbt} & \multicolumn{1}{c|}{0.78} & \multicolumn{1}{c|}{-17.65\%} & -18.54\% & \multicolumn{1}{c|}{1.01} & \multicolumn{1}{c|}{-27.25\%} & -8.60\% \\
 & \texttt{Anchor} & \multicolumn{1}{c|}{0.67} & \multicolumn{1}{c|}{-12.47\%} & -8.97\% & \multicolumn{1}{c|}{0.79} & \multicolumn{1}{c|}{\textbf{-30.32\%}} & \textbf{-10.16\%} \\
 & \texttt{Attn} & \multicolumn{1}{c|}{0.67} & \multicolumn{1}{c|}{-12.12\%} & -8.54\% & \multicolumn{1}{c|}{0.77} & \multicolumn{1}{c|}{-20.22\%} & -6.16\% \\ \hline
\multirow{6}{*}{\textbf{PGD}} & \texttt{Fac} & \multicolumn{1}{c|}{0.97} & \multicolumn{1}{c|}{-10.92\%} & -5.81\% & \multicolumn{1}{c|}{1.6} & \multicolumn{1}{c|}{-35.71\%} & -6.20\% \\
 & \texttt{Mean} & \multicolumn{1}{c|}{0.78} & \multicolumn{1}{c|}{-11\%} & -7.62\% & \multicolumn{1}{c|}{1.92} & \multicolumn{1}{c|}{-35.02\%} & -5.35\% \\
 & \texttt{Hyper} & \multicolumn{1}{c|}{0.79} & \multicolumn{1}{c|}{-10.84\%} & -7.74\% & \multicolumn{1}{c|}{1.8} & \multicolumn{1}{c|}{-34.30\%} & -5.67\% \\
 & \texttt{Mbt} & \multicolumn{1}{c|}{0.72} & \multicolumn{1}{c|}{-10.71\%} & -8.57\% & \multicolumn{1}{c|}{2.58} & \multicolumn{1}{c|}{-53.50\%} & -14.02\% \\
 & \texttt{Anchor} & \multicolumn{1}{c|}{0.87} & \multicolumn{1}{c|}{-12.25\%} & -6.57\% & \multicolumn{1}{c|}{2.52} & \multicolumn{1}{c|}{\textbf{-61.55\%}} & \textbf{-28.41\%} \\
 & \texttt{Attn} & \multicolumn{1}{c|}{0.84} & \multicolumn{1}{c|}{-11.51\%} & -6.10\% & \multicolumn{1}{c|}{1.09} & \multicolumn{1}{c|}{-21.03\%} & -7.43\% \\ \hline
\end{tabular}%
}
\begin{tablenotes}
    \ninept
    \item[a]$^*$quality = 5 for \texttt{Anchor} and \texttt{Attn}
\end{tablenotes}
\end{table*}

\subsection{PGD Attack}
\label{pgd}
PGD~\cite{PGD} is also a powerful and widely used adversarial attack method, which iteratively adjusts the perturbations based on the same calculation formula as FGSM.

In comparison with Algorithm~\ref{fgsm algorithm}, the difference between PGD and iterative FGSM lies in two aspects. 
First, we introduce random noise to the original image before turning it into optimization iterations to raise the robustness and efficiency of the attack. 
Second, the way of constraining total perturbation in PGD is to clamp the noise every iteration. 

Moreover, Madry et al.~\cite{PGD} demonstrate through experiments that the loss value of PGD attack increases fairly consistently and gradually converges, untouched by various start points. 
Thus, the step of iteration in Algorithm~\ref{pgd algorithm} is settled as 40, while the step size is 0.01. 
Noticed that the max perturbation is 0.03 corresponding to the PSNR of around 30, similar with that in FGSM attack.

\begin{algorithm}
    \caption{PGD attack}
    \label{pgd algorithm}
    \begin{algorithmic}[1]
        \REQUIRE original image $X$, iteration step $T$, step size $\delta$, max perturbation size $\epsilon$
        \ENSURE adversarial image $X'$
        \STATE Add random noise $\alpha$: $X'=X+\alpha$
        \STATE Let $t=0$
        \WHILE{$t<T$}
	  \STATE Calculate attack loss L by~\eqref{eq2} or~\eqref{eq3}
	  \STATE Calculate the gradient of the attack loss:\\$grad=\frac{\partial{L(F(X'), X)}}{\partial{X}}$
	  \STATE Obtain noise constrained by max perturbation size: $noise=clamp(X'+\delta{\cdot}sign(grad)-X,-\epsilon,\epsilon)$
        \STATE Obtain the adversarial image: $X'=X+noise$
	  \STATE $t=t+1$
        \ENDWHILE
        \RETURN $X'$
    \end{algorithmic}
\end{algorithm}

\section{Experiment and Evaluations}
\label{experiments}

\subsection{Experiment Setup}
\label{experiment setup}
We choose 6 LIC models mentioned in Section~\ref{learned image comp}, which have already been trained and published in CompressAI~\cite{compressai}. 
CompressAI provides 8 quality levels for pre-trained models. For low quality, we test quality = 2 corresponding to a low bit rate and reconstruction quality. For high quality, we test quality = 5 for \texttt{Anchor} and \texttt{Attn} and quality = 8 for the others.
Iterative FGSM and PGD discussed in Section~\ref{attacks} are universal to all these models.

For the convenience of analyzing experiment results, we define the following parameters:
\begin{equation}
\begin{split}
    bpp~change&=bpp(\widehat{x'})/bpp(\widehat{x})\\
    PSNR~change&=\frac{PSNR(x,\widehat{x'})-PSNR(x,\widehat{x})}{PSNR(x,\widehat{x})}\label{eq4}
\end{split}
\end{equation}

Here $\widehat{x}$ refers to the reconstruction of x. 
As the name implies, bpp change describes the impact on the bit rate. 
However, it should be noted that in PSNR change, the PSNR of $\widehat{x'}$ is calculated with x instead of x', since what we are concerned with is widening the distance between the adversarial reconstruction and the original image.

Since gradients are necessary for both FGSM and PGD to generate adversarial images, we record the gradients of original and adversarial images separately and draw heatmaps of these gradients for visualization.

\begin{figure}[htb]
    \centering
    \subfigure[original image]{
    \label{fig. ori image}
    \includegraphics[width=0.3\linewidth]{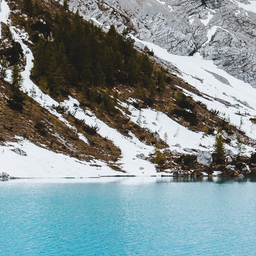}}
    \hspace{0.1em}
    \subfigure[FGSM adversary]{
    \label{fig. fgsm adv}
    \includegraphics[width=0.3\linewidth]{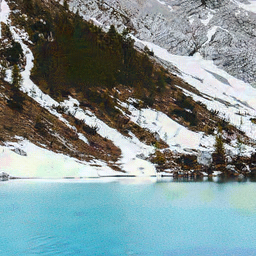}}
    \hspace{0.1em}
    \subfigure[PGD adversary]{
    \label{fig. pgd adv}
    \includegraphics[width=0.3\linewidth]{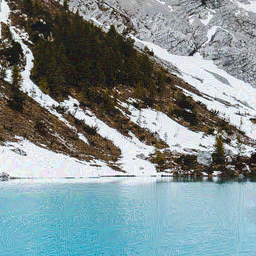}}
    \subfigure[original rebuilding]{
    \label{fig. ori reconstruction}
    \includegraphics[width=0.3\linewidth]{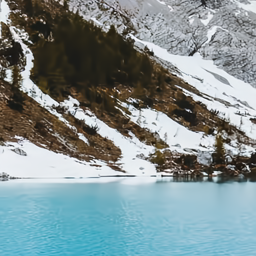}}
    \hspace{0.1em}
    \subfigure[FGSM rebuilding]{
    \label{fig. fgsm reconstruction}
    \includegraphics[width=0.3\linewidth]{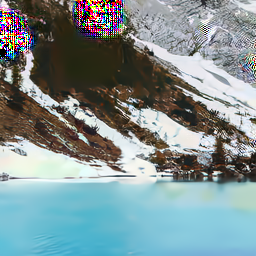}}
    \hspace{0.1em}
    \subfigure[PGD rebuilding]{
    \label{fig. pgd reconstruction}
    \includegraphics[width=0.3\linewidth]{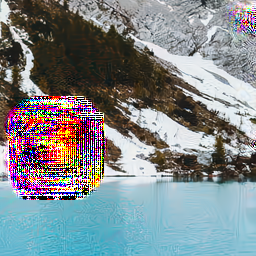}}
    \caption{The results of the FGSM attack (the second column) and the PGD attack (the third column) targeting reconstruction quality on high-quality \texttt{Anchor}.}
    \label{fig. attack}
\end{figure}

\begin{figure}[htb]
    \centering
    \subfigure[original grade]{
    \label{fig. ori grade}
    \includegraphics[width=0.31\linewidth]{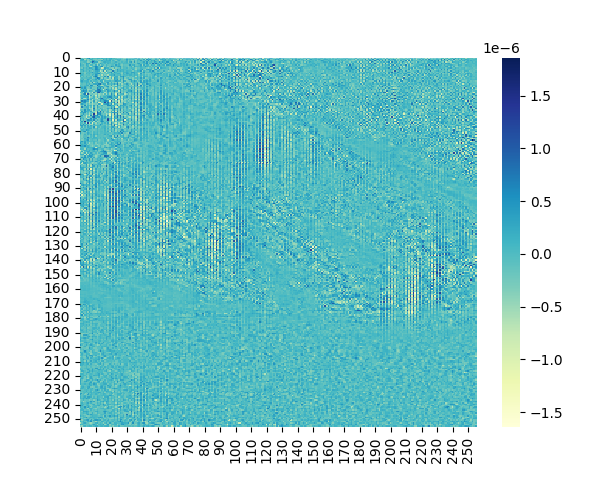}}
    \subfigure[FGSM grade]{
    \label{fig. fgsm grade}
    \includegraphics[width=0.31\linewidth]{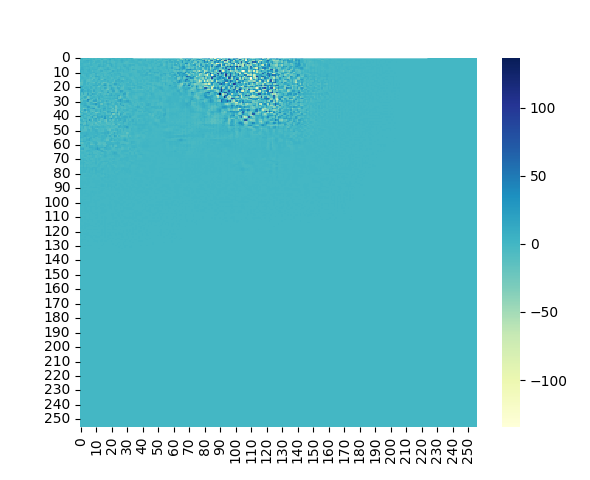}}
    \subfigure[PGD grade]{
    \label{fig. pgd grade}
    \includegraphics[width=0.31\linewidth]{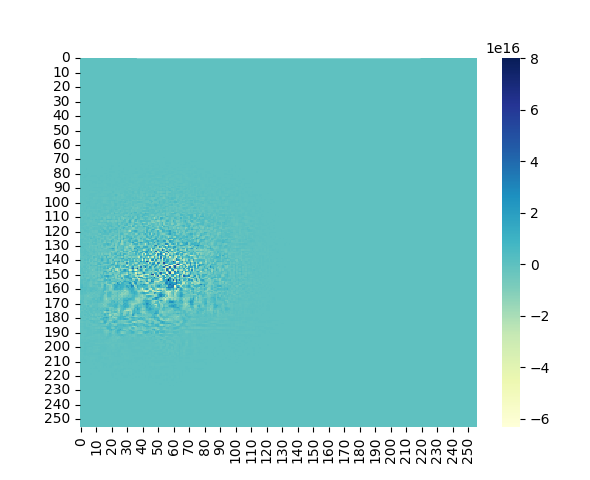}}
    \caption{Gradient heatmaps of original and adversarial images, corresponding to Fig.~\ref{fig. ori image}, \ref{fig. fgsm adv} and \ref{fig. pgd adv}.}
    \label{fig. attack bpp}
\end{figure}

\subsection{Experiment Results}
\label{experiment results}

In Fig.~\ref{fig. ori grade}, it is easy to observe that abnormal values are usually located in the areas of complicated graphics, which vary from different models and qualities.

While targeting the reconstruction quality, abnormal values are concentrated into small areas after being attacked, which embodies the weakness adversarial attack chooses. 
Fig.~\ref{fig. fgsm reconstruction} and Fig.~\ref{fig. pgd reconstruction} show the adversarial reconstructions through the LIC model, whose distortions appear in almost the same positions as these abnormal values in Fig.~\ref{fig. fgsm grade} and Fig.~\ref{fig. pgd grade}.

\begin{table*}[]
\tiny
\centering
\renewcommand{\arraystretch}{1.1}
\caption{Summary of bpp, PSNR, and MS-SSIM change of 6 LIC models under FGSM and PGD attack targeting bit rate.}
\label{bpp attack data}
\resizebox{\textwidth}{!}{%
\begin{tabular}{c|c|ccc|ccc}
\hline
\multirow{2}{*}{Method} & \multirow{2}{*}{Model} & \multicolumn{3}{c|}{quality = 2} & \multicolumn{3}{c}{*quality = 8} \\ \cline{3-8} 
 &  & \multicolumn{1}{c|}{bpp change} & \multicolumn{1}{c|}{PSNR change} & MS-SSIM change & \multicolumn{1}{c|}{bpp change} & \multicolumn{1}{c|}{PSNR change} & MS-SSIM change \\ \hline
\multirow{6}{*}{\textbf{FGSM}} & \texttt{Fac} & \multicolumn{1}{c|}{1.75} & \multicolumn{1}{c|}{-5.23\%} & -3.18\% & \multicolumn{1}{c|}{2.61} & \multicolumn{1}{c|}{-22.80\%} & -3.58\% \\
 & \texttt{Mean} & \multicolumn{1}{c|}{2.79} & \multicolumn{1}{c|}{-1.69\%} & -0.29\% & \multicolumn{1}{c|}{16.32} & \multicolumn{1}{c|}{-26.80\%} & -5.60\% \\
 & \texttt{Hyper} & \multicolumn{1}{c|}{3.28} & \multicolumn{1}{c|}{-2.71\%} & -0.89\% & \multicolumn{1}{c|}{6.59} & \multicolumn{1}{c|}{-29.66\%} & -5.72\% \\
 & \texttt{Mbt} & \multicolumn{1}{c|}{\textbf{18.95}} & \multicolumn{1}{c|}{-1.69\%} & -0.19\% & \multicolumn{1}{c|}{11} & \multicolumn{1}{c|}{-28.54\%} & -5.97\% \\
 & \texttt{Anchor} & \multicolumn{1}{c|}{3.57} & \multicolumn{1}{c|}{-5.16\%} & -1.93\% & \multicolumn{1}{c|}{10.79} & \multicolumn{1}{c|}{-24.51\%} & -7.35\% \\
 & \texttt{Attn} & \multicolumn{1}{c|}{2.35} & \multicolumn{1}{c|}{-4.51\%} & -2.52\% & \multicolumn{1}{c|}{2.94} & \multicolumn{1}{c|}{-15.75\%} & -5.46\% \\ \hline
\multirow{6}{*}{\textbf{PGD}} & \texttt{Fac} & \multicolumn{1}{c|}{1.58} & \multicolumn{1}{c|}{-1.96\%} & -1.55\% & \multicolumn{1}{c|}{2.44} & \multicolumn{1}{c|}{-17.72\%} & -2.21\% \\
 & \texttt{Mean} & \multicolumn{1}{c|}{2.48} & \multicolumn{1}{c|}{0.23\%} & -0.03\% & \multicolumn{1}{c|}{17.73} & \multicolumn{1}{c|}{-20.20\%} & -2.35\% \\
 & \texttt{Hyper} & \multicolumn{1}{c|}{2.62} & \multicolumn{1}{c|}{1.04\%} & 0.56\% & \multicolumn{1}{c|}{5.34} & \multicolumn{1}{c|}{-17.28\%} & -2.02\% \\
 & \texttt{Mbt} & \multicolumn{1}{c|}{\textbf{19.15}} & \multicolumn{1}{c|}{0.85\%} & 0.51\% & \multicolumn{1}{c|}{10.6} & \multicolumn{1}{c|}{-19.68\%} & -2.34\% \\
 & \texttt{Anchor} & \multicolumn{1}{c|}{3.04} & \multicolumn{1}{c|}{-0.69\%} & -0.63\% & \multicolumn{1}{c|}{14.79} & \multicolumn{1}{c|}{-10.85\%} & -2.81\% \\
 & \texttt{Attn} & \multicolumn{1}{c|}{2.27} & \multicolumn{1}{c|}{-1.09\%} & -1.86\% & \multicolumn{1}{c|}{3.73} & \multicolumn{1}{c|}{-8.40\%} & -3.04\% \\ \hline
\end{tabular}%
}
\begin{tablenotes}
    \ninept
    \item[a]$^*$quality = 5 for \texttt{Anchor} and \texttt{Attn}
\end{tablenotes}
\end{table*}

The top half of Table~\ref{psnr attack data} and Table~\ref{bpp attack data} summarizes the results of FGSM attack.  
Among all the conditions, the attack on high-quality \texttt{Anchor} reaches the greatest distortion loss of 30.32\% change while the attack on low-quality \texttt{Mbt} creates the highest increase of bpp by 18.95 times.

We notice that the bpp is saved while attacking the reconstruction quality, especially on low-quality models. 
This is probably due to the theory of FGSM that partly smooths the gradients on the original graphic boundaries to lower the entropy estimation. 
There are also certain losses in PSNR and MS-SSIM while attacking bit rate, which better reflects the effectiveness of our iterative FGSM.

The results of PGD attack are shown in the bottom half of Table~\ref{psnr attack data} and Table~\ref{bpp attack data}, whose worst cases have just the same conclusions as those in the FGSM attack.
While attacking reconstruction quality on high-quality models, the bpp increases, for the reason that PGD introduces greater perturbation than FGSM, resulting in a longer bit length to express the adversarial image. 
Meanwhile, PSNR and MS-SSIM get better especially when attacking bit rate on low-quality models, rather than becoming worse in FGSM, due to the stability of PGD.

\subsection{Discussion}
\label{discussion}
\noindent\textbf{Attack methods.} Compared with FGSM, adding random noise initially improves the error-tolerant rate of PGD so that it can take a larger step size and fewer iterations to reach the convergence, and it is also one-tenth less time-consuming than FGSM.

The MS-SSIM change of high-quality \texttt{Anchor} in PGD attack is greater than that in FGSM attack because MS-SSIM emphasizes human perception. 
It can be confirmed by their reconstructions, as shown in Fig~\ref{fig. fgsm reconstruction} and Fig~\ref{fig. pgd reconstruction}, more visible distortions exist in the reconstruction after the PGD attack. 
Meanwhile, abnormal values over the gradient heatmaps in the PGD attack also appear more conspicuous if we compare  Fig.~\ref{fig. fgsm grade} with Fig.~\ref{fig. pgd grade}.

However, iterative FGSM shows its advantage in attacking low-quality models, perhaps because the smaller step size makes iterative FGSM fine to discover weaknesses in low-bit-rate images.

\noindent\textbf{Models.} For the PSNR attack, \texttt{Anchor} using residual blocks with small convolutions is the most vulnerable while the PSNR change of \texttt{Attn} is only one-third of the former. 
For the bpp attack, \texttt{Fac} is the most robust despite its simplicity. 
Low-quality \texttt{Mbt} with autoregressive context model makes the greatest loss in bit rate.

In general, \texttt{Attn} with self-attention block shows the best robustness in all the conditions, which means the self-attention block contributes to comprehending the original information of adversarial images. 

\noindent\textbf{Qualities.} The PSNR change of high-quality models is more dramatic since high bpp retains more information including noise during the compression process. 
The results of the bpp attack follow the above regularities except that low-quality \texttt{Mbt} is more vulnerable than the high-quality one.

\subsection{Attack on Conventional Methods}
\label{conventional methods}
Conventional methods use reversible transform to compress image information, which is simpler than LIC models. 
Generally speaking, gradient-based methods are invalid for conventional methods. 
Thus, take H.266 as an example, we consider if the high transferability of one-step FGSM discovered by Papernot et al.~\cite{transfer} and Dong et al.~\cite{tranferfgsm} is effective in transferring attack to H.266.

In the black-box setting, we adopt one-step FGSM on the most vulnerable LIC model to generate adversarial images and use VTM to encode them, while our VTM encoder version is 19.0.
Compared with high-quality \texttt{Anchor}, we set the quantization parameter Q=20 in the PSNR attack. 
Similarly, Q is set as 40 in the bpp attack. 
The overall coding results are shown in Table~\ref{conventional attack data}.

\begin{table}[t]
	\centering
	\renewcommand{\arraystretch}{1.1}
	\caption{Results of transferring attack on H.266 targeting reconstruction quality and bit rate}
	\label{conventional attack data}
	\resizebox{\columnwidth}{!}{%
		\begin{tabular}{c|cccc|cccc}
			\hline
			& \multicolumn{4}{c|}{\textbf{PSNR attack}} & \multicolumn{4}{c}{\textbf{bpp attack}} \\ \cline{2-9} 
			& \multicolumn{2}{c|}{\begin{tabular}[c]{@{}c@{}}\texttt{Anchor}\\ (quality=5)\end{tabular}} & \multicolumn{2}{c|}{H.266 (Q=20)} & \multicolumn{2}{c|}{\begin{tabular}[c]{@{}c@{}}\texttt{Mbt}\\ (quality=2)\end{tabular}} & \multicolumn{2}{c}{H.266 (Q=40)} \\ \cline{2-9} 
			\multirow{-3}{*}{} & \multicolumn{1}{c|}{PSNR} & \multicolumn{1}{c|}{bpp} & \multicolumn{1}{c|}{PSNR} & \multicolumn{1}{c|}{bpp} & \multicolumn{1}{c|}{PSNR} & \multicolumn{1}{c|}{bpp} & \multicolumn{1}{c|}{PSNR} & \multicolumn{1}{c}{bpp} \\ \hline
			\multicolumn{1}{l|}{Origin} & \multicolumn{1}{c|}{35.12} & \multicolumn{1}{c|}{0.59} & \multicolumn{1}{c|}{46} & {\color[HTML]{333333} 0.66} & \multicolumn{1}{c|}{{\color[HTML]{333333} 29.64}} & \multicolumn{1}{c|}{{\color[HTML]{333333} 0.19}} & \multicolumn{1}{c|}{{\color[HTML]{333333} 34.44}} & {\color[HTML]{333333} 0.04} \\ \hline
			\multicolumn{1}{l|}{Adversary} & \multicolumn{1}{c|}{8.52} & \multicolumn{1}{c|}{1.55} & \multicolumn{1}{c|}{43.58} & {\color[HTML]{333333} 2.4} & \multicolumn{1}{c|}{{\color[HTML]{333333} 23.6}} & \multicolumn{1}{c|}{{\color[HTML]{333333} 0.56}} & \multicolumn{1}{c|}{{\color[HTML]{333333} 26.97}} & {\color[HTML]{333333} 0.15} \\ \hline
		\end{tabular}%
	}
\end{table}

As for the PSNR attack, H.266 performs much better than LIC models, since frequency domain transform smooths the high-frequency noise FGSM adds. 
However, the cost is that the bpp of H.266 also raises in the PSNR attack, as the high-frequency noise is more difficult to compress and occupies a longer bit length. 
That`s why H.266 also shows its vulnerability in the bpp attack.
Even so, conventional methods can probably be used as a defense measure against the PSNR attack if higher bpp is allowable.

\begin{table*}[]
\centering
\renewcommand{\arraystretch}{1.1}
\caption{Finetuning results of adversarial training against PSNR attack and bpp attack}
\label{finetune data}
\resizebox{\textwidth}{!}{%
\begin{tabular}{l|cccc|cccc}
\hline
\multicolumn{1}{c|}{} & \multicolumn{4}{c|}{\textbf{Denfense against PSNR attack}} & \multicolumn{4}{c}{\textbf{Denfense against bpp attack}} \\ \cline{2-9} 
\multicolumn{1}{c|}{\multirow{-2}{*}{}} & \multicolumn{1}{c|}{\begin{tabular}[c]{@{}c@{}}PSNR change\\ (attacked)\end{tabular}} & \multicolumn{1}{c|}{\begin{tabular}[c]{@{}c@{}}R-D cost\\ on original images\end{tabular}} & \multicolumn{1}{c|}{\begin{tabular}[c]{@{}c@{}}R-D cost on\\ adversarial images\end{tabular}} & \multicolumn{1}{c|}{\begin{tabular}[c]{@{}c@{}}R-D cost loss\\ (\%)\end{tabular}} & \multicolumn{1}{c|}{\begin{tabular}[c]{@{}c@{}}bpp change\\ (attacked)\end{tabular}} & \multicolumn{1}{c|}{\begin{tabular}[c]{@{}c@{}}R-D cost\\ on original images\end{tabular}} & \multicolumn{1}{c|}{\begin{tabular}[c]{@{}c@{}}R-D cost on\\ adversarial images\end{tabular}} & \multicolumn{1}{c}{\begin{tabular}[c]{@{}c@{}}R-D cost loss\\ (\%)\end{tabular}} \\ \hline
Pre-trained model & \multicolumn{1}{c|}{-61.55\%} & \multicolumn{1}{c|}{1.095} & \multicolumn{1}{c|}{74.05} & 6662.00\% & \multicolumn{1}{c|}{19.15} & \multicolumn{1}{c|}{0.434} & \multicolumn{1}{c|}{3.813} & 778.00\% \\ \hline
Finetuned model & \multicolumn{1}{c|}{-12.28\%} & \multicolumn{1}{c|}{1.69} & \multicolumn{1}{c|}{3.316} & {\color[HTML]{333333} 96.21\%} & \multicolumn{1}{c|}{{\color[HTML]{333333} 1.44}} & \multicolumn{1}{c|}{{\color[HTML]{333333} 0.583}} & \multicolumn{1}{c|}{{\color[HTML]{333333} 0.556}} & {\color[HTML]{333333} -4.63\%} \\ \hline
Finetuning effect$^*$ & \multicolumn{1}{c|}{-} & \multicolumn{1}{c|}{54.34\%} & \multicolumn{1}{c|}{\textbf{-95.52\%}} & \multicolumn{1}{c|}{{\color[HTML]{333333} -}} & \multicolumn{1}{c|}{{\color[HTML]{333333} -}} & \multicolumn{1}{c|}{{\color[HTML]{333333} 34.33\%}} & \multicolumn{1}{c|}{{\color[HTML]{333333} \textbf{-85.42\%}}} & \multicolumn{1}{c}{{\color[HTML]{333333} -}} \\ \hline
\end{tabular}%
}
\begin{tablenotes}
    \item[a] $^*$Finetuning effect is represented by the R-D cost change, which is calculated similarly to the PSNR change as in~\eqref{eq4}.
\end{tablenotes}
\end{table*}

\section{Defense Method}
\label{defense method}

\subsection{Adversarial Training}
\label{adversarial training}
Known as PGD training, the defense method proposed by Madry et al.~\cite{PGD} is one of the most effective ways to weaken the misdirection of neural networks caused by adversarial attacks. 
PGD training uses adversarial images generated by PGD to train the network. 
Thus, except for its effectiveness, PGD training requires numerous computing resources and high time complexity (relevant to the iteration steps of the PGD attack).

However, Madry et al. designed this method for image classification tasks rather than compression. 
The difference is embodied in that classification tasks do not pursue reconstruction quality, but only demand recognition accuracy. In contrast, a good reconstruction must be close enough to the original image for image compression. 
Hence, the finetuned LIC models should learn the features from not only adversarial but also original images. 
We modify the algorithm by adding adversarial images into the training dataset instead of replacing it, as shown in Algorithm~\ref{training algorithm}. 
The finetuning goal can be changed by the target of the PGD attack (line 4 in Algorithm~\ref{training algorithm}).

\begin{algorithm}
    \caption{PGD training}
    \label{training algorithm}
    \begin{algorithmic}[1]
        \REQUIRE Randomly initialized network F, original training set X
        \REPEAT
        \STATE Read minibatch $B=\{x^1,\cdots,x^m\}$ from X
	  \STATE Train one step of F with minibatch B
	  \STATE Generate adversarial minibatch $B'=\{x^1_{adv},\cdots,x^m_{adv}\}$ by PGD attack
	  \STATE Train one step of F with minibatch B'
        \STATE Valid F with both minibatch B and B'
        \UNTIL training convergence
    \end{algorithmic}
\end{algorithm}

Our training set uses randomly cropped 256*256*3 patches of 1633 images from the clic2020 dataset~\cite{clic2020} as the inputs, with a train batch size of 16. 
The $\lambda$ in R-D cost is the same as the corresponding pre-trained models in CompressAI. 
The initial learning rate is set as 0.0002 while the max epoch is 200.

\subsection{Defense Efficiency}
\label{defense efficiency}
To evaluate the effect of our modified defense method, we attack pre-trained and finetuned models with PGD and compare their R-D costs. 
Settings of the PGD attack are the same as in Sec.~\ref{pgd}. 
Although PGD training is specifically designed for one finetuning goal, sacrificing image quality to obtain lower bpp is unacceptable.

As shown in Table~\ref{finetune data}, the horizontal comparison describes the loss brought by adversarial images while the vertical comparison indicates the effect of model finetuning.

We choose high-quality \texttt{Anchor} as the target for defending against the PSNR attack because it is the most vulnerable model while attacking the reconstruction quality. 
Table~\ref{finetune data} shows that the loss from adversarial images is greatly weakened. 
Furthermore, the R-D cost of the finetuned model on adversarial images decreases by 95.52\%. 
Similarly, we use low-quality \texttt{Mbt} to defend against the bpp attack. 
The fine-tuned model reached an 85.42\% R-D cost decrease on adversarial images, even lower than the R-D cost on original images as the PGD attack raises the PSNR of reconstructions.
However, the R-D cost of the finetuned model increases on original images compared to the pre-trained model, which is inevitable due to the `polluted’ dataset in the training process.

Since the PGD attack is to find the most adversarial image in the neighbor ball around the original image, our experiment results prove that the modified algorithm is effective in defending both one-step and iterative adversarial attacks, whether targeting for reconstruction quality or bit rate. 
This method could be generalized to most LIC models because of its successful defense of the most vulnerable ones in our experiment.

\section{Conclusion}
\label{conclusion}
In this paper, we research the robustness of learned image compression networks under gradient-based adversarial attack methods. 
We demonstrate the decrease of compression ability against either PSNR or bpp attacks. 
We also test the robustness of H.266, which prompts future research that conventional methods could be used to defend against adversarial attacks.  
Furthermore, we supplement the defense method of adversarial training to enhance the performance of LIC models under attack.

\bibliographystyle{IEEEbib}
\bibliography{refs}

\end{document}